\documentclass{IEEEtran4PSCC}
\ifCLASSINFOpdf
   \usepackage[pdftex]{graphicx}
\else
   \usepackage[dvips]{graphicx}
\fi
\usepackage{hyperref}

\usepackage{svg}
\usepackage{graphicx}
\usepackage{subcaption}

\usepackage{multirow}
\makeatletter
\let\old@ps@headings\ps@headings
\let\old@ps@IEEEtitlepagestyle\ps@IEEEtitlepagestyle
\def\psccfooter#1{%
    \def\ps@headings{%
        \old@ps@headings%
        \def\@oddfoot{\strut\hfill#1\hfill\strut}%
        \def\@evenfoot{\strut\hfill#1\hfill\strut}%
    }%
    \def\ps@IEEEtitlepagestyle{%
        \old@ps@IEEEtitlepagestyle%
        \def\@oddfoot{\strut\hfill#1\hfill\strut}%
        \def\@evenfoot{\strut\hfill#1\hfill\strut}%
    }%
    \ps@headings%
}
\makeatother

\psccfooter{%
        \parbox{\textwidth}{\hrulefill \\ \small{22nd Power Systems Computation Conference} \hfill \begin{minipage}{0.2\textwidth}\centering \vspace*{4pt} \includegraphics[scale=0.06]{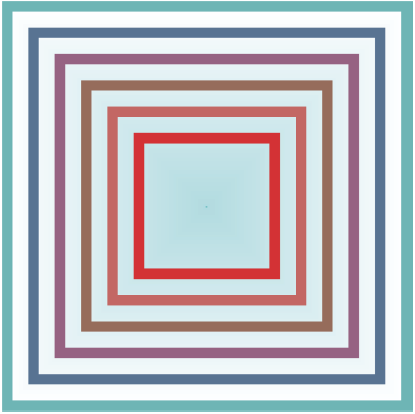}\\\small{PSCC 2022} \end{minipage} \hfill \small{Porto, Portugal --- June 27 -- July 1, 2022}}%
}

\usepackage{amsmath,amssymb,amsfonts}
\usepackage{graphicx}
\usepackage{textcomp}
\usepackage{xspace}
\usepackage[T1]{fontenc}
\usepackage{bm}
\usepackage{algorithmic}

\usepackage{float}
\usepackage[font=small]{caption}

\floatstyle{ruled}
\newfloat{model}{t!}{log}
\floatname{model}{\textsc{Model}}
\newfloat{problem}{t!}{log}
\floatname{problem}{\textsc{Problem}}

\newcommand{\LP}{\bm{{Simp-LP}}\xspace} %  simplified LP 
\newcommand{\ExtLP}{\bm{{Extn-LP}}\xspace} %  extended LP 
\newcommand{\Exc}{\bm{{Exact-MILP}}\xspace} %  Exact formulation
\newcommand{\NA}{\bm{{NaAl-LP}}\xspace} %  Nazir-Almassalkhi formulation formulation

\newcommand{\Relyz}{\bm{{Relax-LP}}\xspace} %  Relaxed MILP formulation

\newcommand{\unif}{\mathcal{U}}

\usepackage{color,soul} % highlights
\begin{document}
\title{Linear Battery Models for Power Systems Analysis}

%% To specify the authors when (number of affiliations <= 2)
\author{
\IEEEauthorblockN{David Pozo}
\IEEEauthorblockA{Center for Energy Science and Technology\\
Skolkovo Institute of Science and Technology (Skoltech)\\
Moscow, Russia \\
}
}

%% To specify the authors when (number of affiliations > 2)
% \author{\IEEEauthorblockN{Author n.1\IEEEauthorrefmark{1},
% Author n.2\IEEEauthorrefmark{2},
% Author n.3\IEEEauthorrefmark{3}, 
% Author n.4\IEEEauthorrefmark{3} and
% Author n.5\IEEEauthorrefmark{4}}
% \IEEEauthorblockA{\IEEEauthorrefmark{1} Department Name of Organization A\\
% Name of the organization A,
% Address A\\ Emails if wanted}
% \IEEEauthorblockA{\IEEEauthorrefmark{2} Department Name of Organization B\\
% Name of the organization B,
% Address B\\ Emails if wanted}
% Name of the organization C,
% Address C\\ Emails if wanted}
% \IEEEauthorblockA{\IEEEauthorrefmark{4}Department Name of Organization D\\
% Name of the organization D,
% Address D\\ Emails if wanted}
% }

\maketitle
\begin{abstract}
Mathematical models are just models. 
The desire to describe battery energy storage system (BESS) operation using computationally tractable model formulations has motivated a long-standing discussion in both the scientific and industrial communities. Linear BESS models are the most widely used so far. However, finding suitable linear BESS models has been controversial.

This paper focuses on the description of linear BESS models. Four linear BESS formulations are presented, among the most popularly used. A new formulation is also proposed. The 5 BESS models are tested in 100 random BESS and 1.450 random samples of daily profiles of renewable generation.  Two classical problems of power systems, namely, the set-point tracking problem and the transmission expansion planning problem, are selected for numerical analysis. Five thousand simulations are used to draw a better interpretation of each linear formulation presented and showcase specific challenges of BESS models. Practical recommendations are provided based on the findings. 
\end{abstract}


\section{Introduction}

% \subsection{Motivation and existing literature}
There is increasing interest in the modeling of battery energy storage systems (BESS) in the power system community due to the key role of such technologies in future power grids \cite{akhil2015doe}.
Although BESS behavior is non-linear, there has been much interest in modeling BESS as a linear set of constraints \cite{sioshansi2021energy}.
As such, the generic and ideal energy storage model \cite{pozo2014unit} is among one of the most used linear model for power system operation and planning analysis. 

Apart from the accuracy issues for using linear models, it is still missing an explicit formulation for accounting for the mutually exclusive operational states, namely charging, discharging, and idle. Additional complementary equations can be imposed at the cost of increasing computational complexity \cite{arroyo2020use}.  In recent years, several works have tried to overcome such limitations by proposing conditions in which complementary values would be satisfied using linear BESS models.  For example, in \cite{zhao2018using} it was mentioned that the binary representation of the complementary nature of the charging, discharging and idle states can be dropped for storage round-trip efficiencies smaller than one in the context of transmission expansion planning problem. In reference \cite{wen2015enhanced} similar conditions were provided when a linear BESS model is used in the unit commitment problem.  However, a recent study proved them wrong by counterexamples \cite{arroyo2020use}.  To date, no previous studies have proposed an alternative linear formulation that satisfies the complementary conditions. The consequences of finding linear formulations (polytime solvable) that satisfy complementarity would devise exceptional outcomes in complexity theory and the P versus NP problem. It would be equivalent to finding a polynomial-time-solvable formulation of an MIP problem.

% \subsection{Paper organization and contributions}
In this work, we examine the most useful linear BESS models and their limitations. It also argues about limitations for finding a new family of linear BESS models. Furthermore, a new linear BESS formulation is introduced. 
Insights on suboptimality and infeasibility issues for 5 BESS model formulation are discussed using an illustrative example and extensive numerical simulations for two classical power systems problems, the set-point tracking problem and the transmission expansion planning. One hundred random instances for each problem are tested in the 5 BESS linear formulations. An extensive set of metrics to assess the performance of each BESS formulation for different types of problems are presented.

The rest of the paper is organized as follows. Section \ref{sec. Linear} introduces, in crescendo, five linear formulations for the operational modeling of BESS. Section \ref{sec. Numerical} describes the numerical experiments for the two problems mentioned. Finally, conclusions and discussion are given in section \ref{sec. conclusions}.

\section{Linear BESS model Formulations}
\label{sec. Linear}

\subsection{General considerations}

In this paper, we focus on modeling an \textit{generic and ideal} energy storage device defined in \cite{pozo2014unit}. It is defined as follows: ``\textit{a \emph{generic storage} device [is] any device with the ability to transform and store energy, and reverse the process by injecting the stored energy back into the system [while] a \emph{ideal storage} device assumes certain simplifications in its technical and economic operation.''}
In this setting, a mathematical model for a generic and ideal  BESS is practical for integration with other large mathematical programming models for applications in power system operation and planning. There are three fundamental elements that characterize BESS mathematical formulations, namely operational constraints, decision variables, and parameters. 

The \textit{parameters} are constants for the mathematical formulation.  Specifically, the parameters are the minimum and maximum energy storage capacity $\underline{E}$ and $\overline{E}$, the efficiency rate of energy storage / production (charge / discharge), $\eta^c < 1$ and $\eta^d < 1$, the maximum charging and discharging power rates $\overline{P}^c$ and $\overline{P}^d$. Other works \cite{gonzalez2019non} and \cite{pandvzic2018accurate} have provided alternative BESS mathematical modeling for the case of nonconstant parameters. It is out of the scope of the analysis of this paper. 

BESS mathematical models require a minimum of two \textit{decision variables}\footnote{There are other works considering one decision variable representing the net BESS power exchanges \cite{nazir2021guaranteeing}. However, it does not allow differentiation between charging and discharging efficiency rates.}, charging and discharging power rates, $p^c$ and $p^d$. However, it is common to use additional decision variables that facilitate the mathematical formulation of operational constraints and their interpretation, such as the state of energy (SoE) $e$, representing the energy level in a BESS.
% charging state $z$, taking value 1 if charging and 0 otherwise, and the discharging state $y$. 

The fundamental BESS model \textit{constraints} should capture $\bm{(i)}$ the power limits of BESS charging and discharging rate, $\bm{(ii)}$  the minimum and maximum limits of the energy stored in the BESS, and $\bm{(iii)}$ the SoE backlog carried from one time period to the next. The latter couples two consecutive periods and is typically the only time-coupling constraint in most of the BESS models. $\bm{(iv)}$ The last relevant constraint, but not always fundamental, is the complementarity condition for not charging and discharging simultaneously, i.e., $p^c p^d =0$.

Hereinafter, two simplifications are assumed when introducing BESS models. Firstly, the time-step duration to transform  ``power'' variables into ``energy'' variables is omitted. BESS models are exact for a time step equals to one hour; in other cases, adding an step interval constant in the energy-related constraints is straightforward.  
Secondly, a single period is used for presenting BESS models. The intertemporal connection with the previous period is made by the variable $E_0$. For descriptive purposes, $E_0$ will be given as a parameter, but the reader should keep in mind that it is a variable in a multiperiod formulation. 
%We also highlight it in the cases that it could be an important issue.  
Both simplifications do not compromise the accuracy of the models presented, while easing the mathematical description.

\subsection{The  most common linear BESS models}

Two main models has been extensively used in the literature of power systems for techno-economic analysis.  The first, an exact formulation, \Exc, described by \eqref{eq. exact}. \Exc accounts for the backlog energy in the battery \eqref{eq. exc energy}, charging and discharging power capacity rates \eqref{eq. Exc PcMax} and \eqref{eq. Exc PdMax}, and the minimum and maximum SoE limits \eqref{eq. exc E_lim}.
In addition, the BESS could either charge or discharge, but not both at the same time. This complementarity condition is modeled using binary variables $z$ and $y$ to represent the charging state ($z = 1, y=0$), the discharging state ($z = 0, y= 1$) or idle time ($z = 0, y=0$). Note that \eqref{eq. No charge_discharge} avoids simultaneous charging and discharging. 
Due to the presence of binary variables, this formulation would lead to mixed integer linear programs (MILP).

%%% Exact formualtion
\begin{model}[ht!]
\caption{Exact Formulation \hfill (\Exc)}
\label{Mod: exact}
\begin{subequations} \label{eq. exact}
\begin{IEEEeqnarray}{rl}
&  e =  E_0 + \eta^c p^c - \frac{1}{\eta^d}p^d  \label{eq. exc energy} \\
&  0 \leq  p^c  \leq \overline{P}^c z  \label{eq. Exc PcMax}\\
&  0 \leq  p^d  \leq \overline{P}^d y \label{eq. Exc PdMax}\\
&  y + z \leq  1   \label{eq. No charge_discharge} \\
&   \underline{E} \leq  e \leq \overline{E}  \label{eq. exc E_lim}
\end{IEEEeqnarray}
\mbox{\bf Variables: } $ e, p^c, p^d \in \mathbb{R}_{\geq 0}, z, y \in \left\{0,1 \right\} $
\end{subequations}
\end{model}

The second common formulation does not avoid simultaneous charging and discharging. Known in this paper as \LP, it is a simplification of the formulation \Exc by eliminating binary variables.   It is probably the most common mathematical description in literature for BESS and other storage-like devices. Its popularity arises from the simplicity of it for capturing the fundamental power rate and energy capacity limits as well as the backlog states. But, more importantly, it is a set of linear constraints. Embedding this set of linear constraints into any operational and planning techno-economic model would impact very little into the increase of computational complexity. 

%%% Relaxed LP formualtion
\begin{model}[h!]
\caption{Simplified formulation \hfill (\LP)}
% \label{Mod: relased LP}
\begin{subequations} \label{eq. relaxed}
\begin{IEEEeqnarray}{rl}
&  e =  E_0 + \eta^c p^c - \frac{1}{\eta^d}p^d   \\
&  0 \leq  p^c  \leq \overline{P}^c  \label{eq. LP PcMax}\\
&  0 \leq  p^d  \leq \overline{P}^d \label{eq. LP PdMax}\\
&   \underline{E} \leq  e \leq \overline{E} 
 \end{IEEEeqnarray}
\mbox{\bf Variables: } $ e, p^c, p^d \in \mathbb{R}_{\geq 0}$
\end{subequations}
\end{model}

A two-period illustrative example is introduced to help the interpretation of the presented BESS models. 

\vspace{0.2cm}

\begin{figure}[bht]
\centering
    \includegraphics[width=1\linewidth]{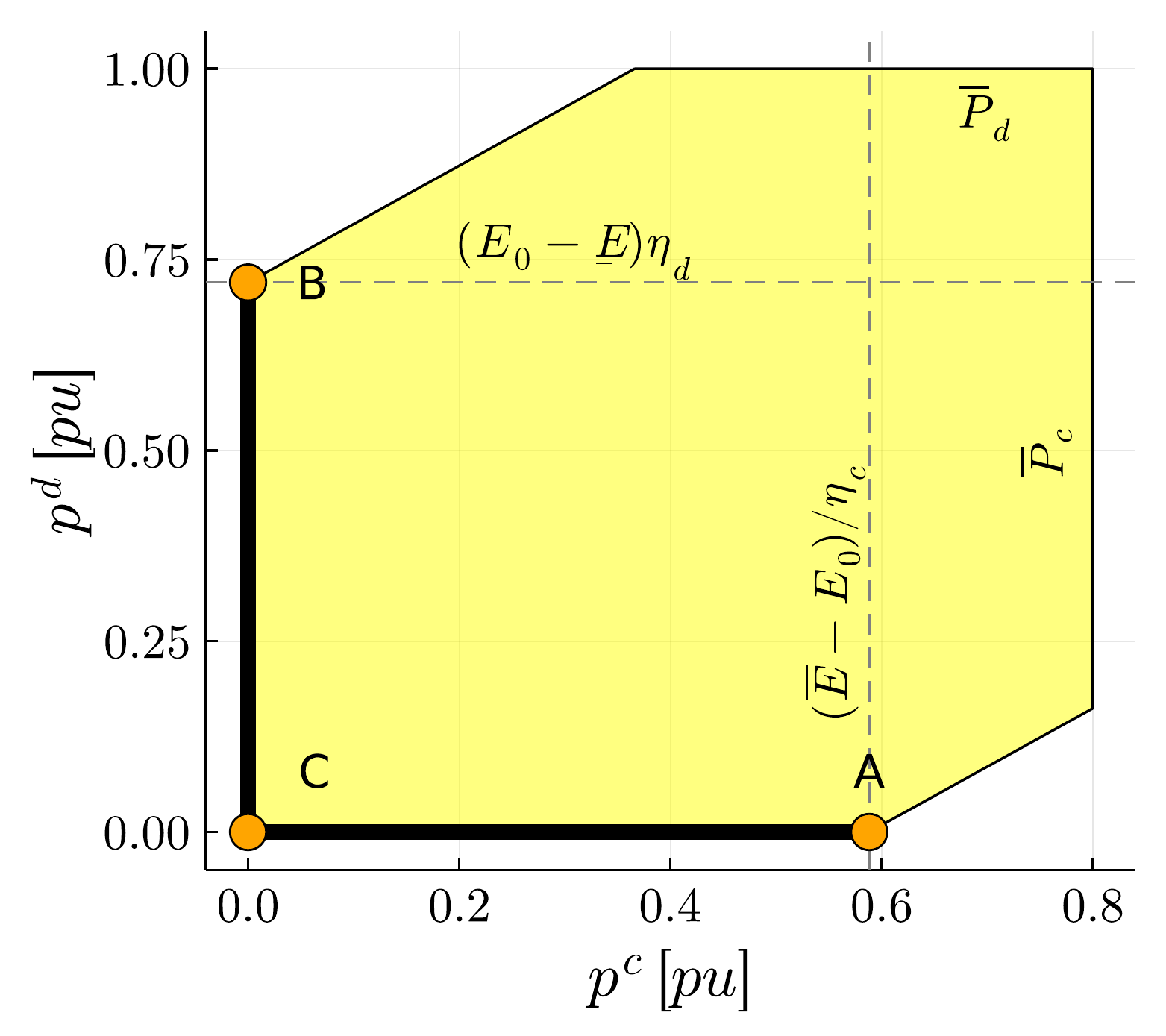}
	\caption{Feasible operating region in  the $(p^c,p^d)$-space for the \Exc model \eqref{eq. exact},  in black, and the \LP  BESS model \eqref{eq. relaxed} in yellow. Orange dots are representative limits of the true feasible operating region. }
	\vspace{0.5cm}
	\label{fig. exact_vs_relaxed}
\end{figure}

\begin{figure*}[hbt]
\centering
	\begin{minipage}{0.32\textwidth}
% 		\vspace{-0.35cm}
    \centering
    \includegraphics[width=1\textwidth]{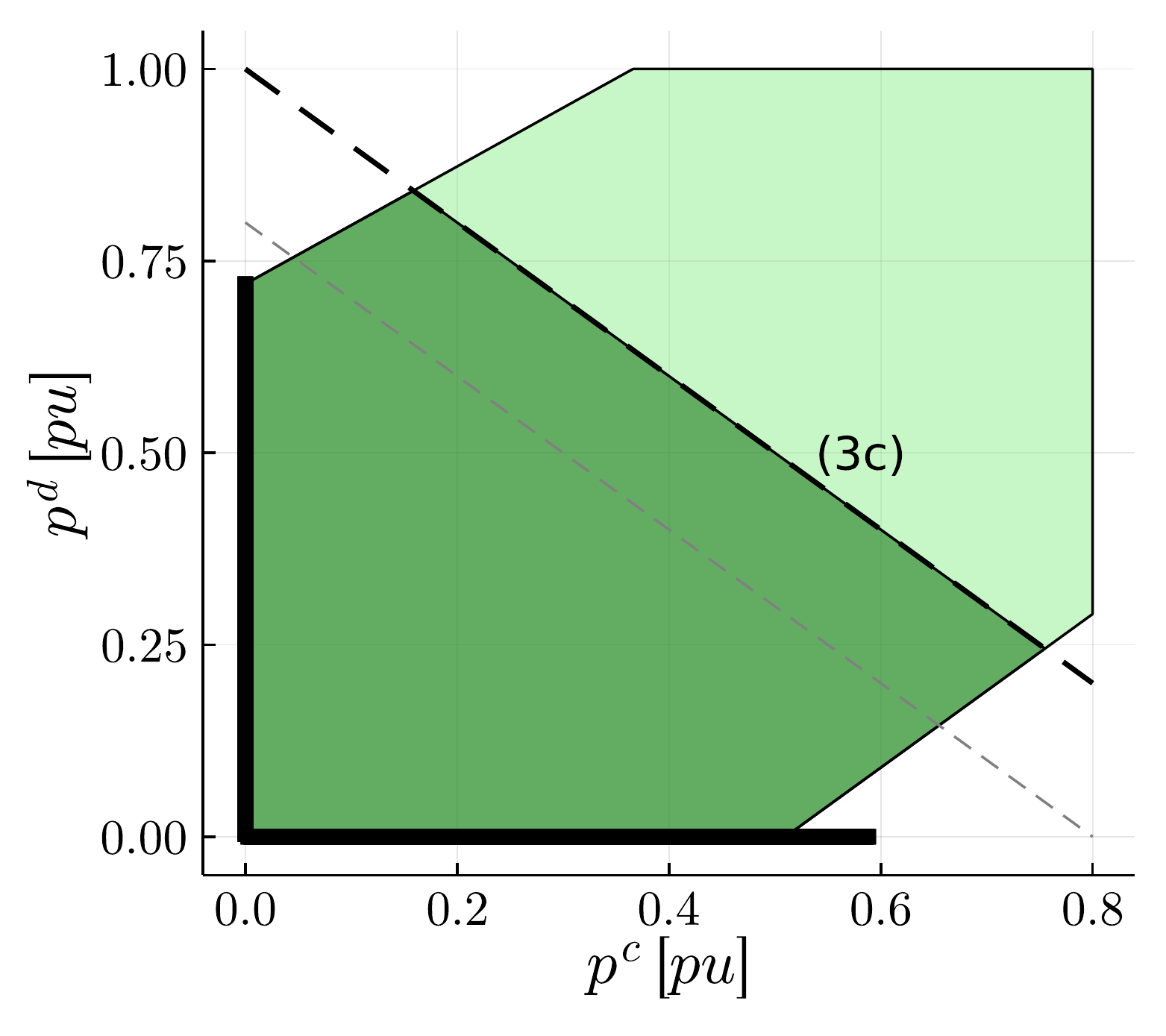}
    \caption*{(a) \NA vs. \Exc}
	\end{minipage}
	\hspace{0.05cm}
	\begin{minipage}{0.32\textwidth}
% 		\vspace{-0.5cm}
    \centering
    	\includegraphics[width=1\linewidth]{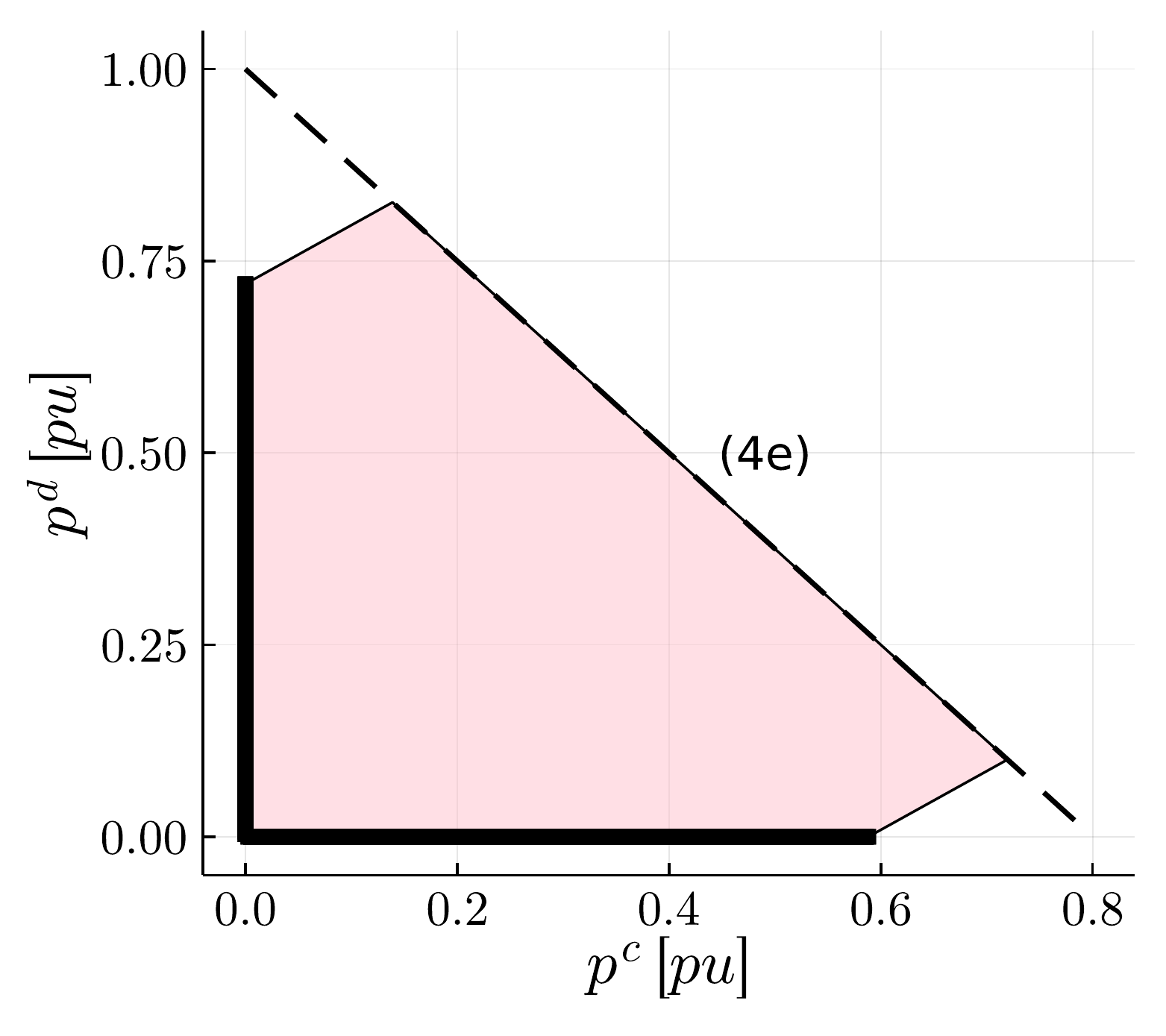}
    	 \caption*{(b) \Relyz vs. \Exc}
	\end{minipage}
	\hspace{0.05cm}
	\begin{minipage}{0.32\textwidth}
% 		\vspace{-0.35cm}
    \centering		\includegraphics[width=1\textwidth]{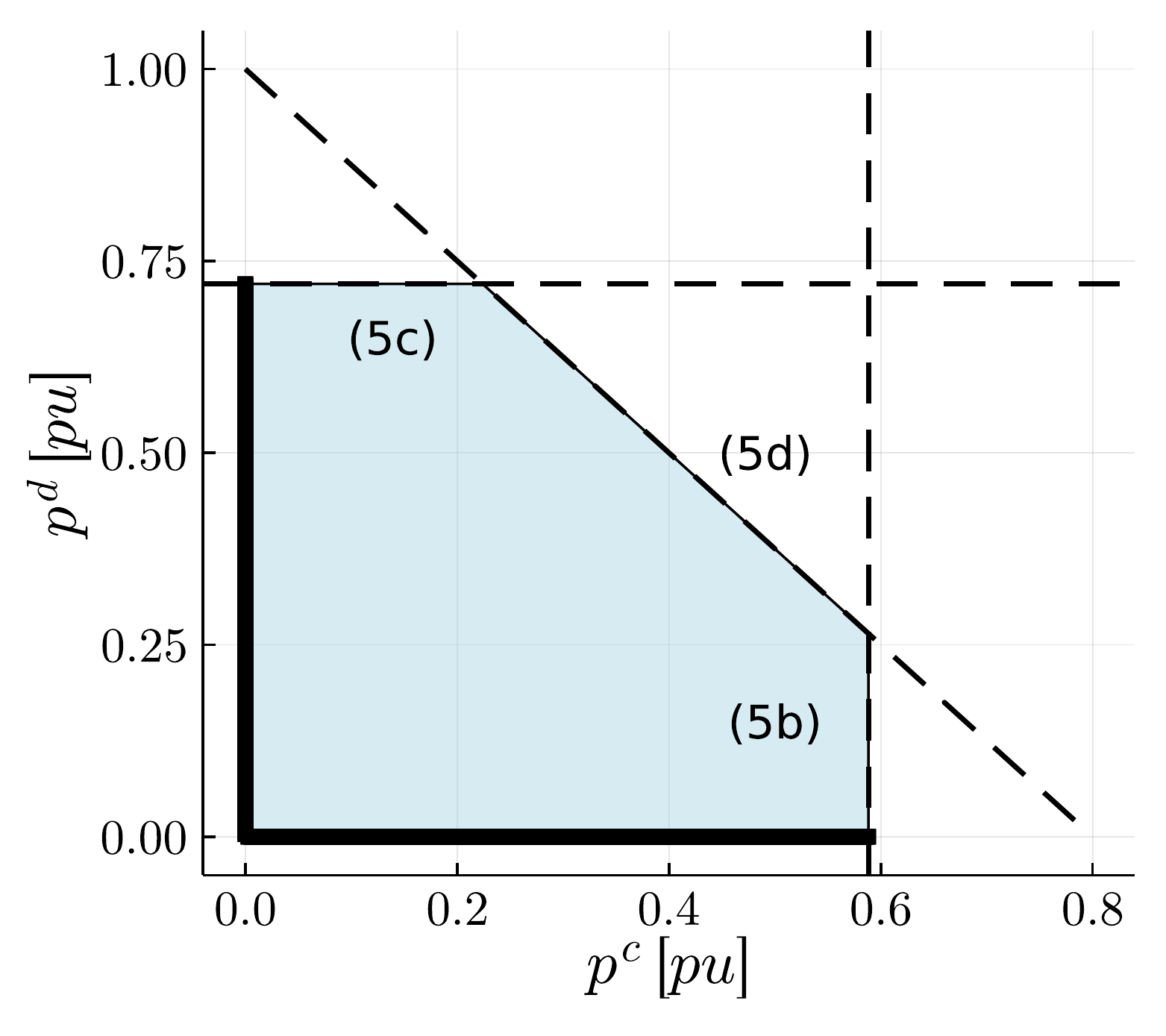}
    \caption*{(c) \ExtLP vs. \Exc}
	\end{minipage}
	\caption{Charging and discharging power feasible region for a single BESS. BESS parameters are in Table \ref{tab.parameters}, corresponding with the example~1.}
% 	\textit{{(Left)}} In dark green the \NA formulation \eqref{eq. Nazir}, and in light green the \NA robust formulation without equation \eqref{eq. Pc + Pd NA}. 
% 	{\textit{(Center)}} \Relyz formulation in red \eqref{eq. relaxed_zy}. 
% 	\textit{{(Right)}} \ExtLP  formulation \eqref{eq. extended_LP} in light blue.}
	\vspace{-0.5cm}
	\label{fig: feasible regions}
	\vspace{0.5cm}
\end{figure*}

\noindent \textbf{Example 1.} \textit{Consider a BESS characterized by the parameters in Table \ref{tab.parameters}. Parameters' units are in pu. 
% Initial SoE is $E_0=0.7$, for period $0$. 
BESS operation feasible regions in  the $(p^c,p^d)$-space are presented in Fig.~\ref{fig. exact_vs_relaxed} and Fig.~\ref{fig: feasible regions}. }

\begin{table}[h!] 
\renewcommand{\arraystretch}{1.3} % adds row cushion
\centering
\caption{Illustrative BESS example parameters}
\begin{tabular}{|c c c|c c|c c|}  
\hline
$\mathbf{\underline{E}}$  & $\mathbf{\overline{E}}$ & $\mathbf{{E}_0}$ & $\mathbf{\overline{P}^c}$  & $\mathbf{\overline{P}^d}$  & $\mathbf{\eta^c}$ & $\mathbf{\eta^d}$ \\
 \hline      
  \hline
0.7 & 2 & 1.5 & 0.8 & 1 & 0.85 & 0.9 \\
\hline
\end{tabular} 
\label{tab.parameters}
\end{table}
    
\Exc and \LP feasible regions for charging and discharging power in the $(p^c,p^d)$-space are represented in Fig.~\ref{fig. exact_vs_relaxed} in a thick black line and a yellow region, respectively. 
% provide the feasible  charging and discharging points from the BESS model \Exc, while the yellow region from the \LP. 

Point A represents the actual maximum charging power rate. It is bounded by its explicit limit $\overline{P}^c$ and the charging power related to the remaining energy capacity for charging $(\overline{E} - E_0)/\eta^c$. 
Thus, in our illustrative example, point A is equal to $\min \left\{0.8,  (2 - 1.5)/0.85 \right\} = 0.588 $. 
Similarly, point B represents the maximum actual discharging power rate. It is bounded by its explicit limit $\overline{P}^d$ and the discharging power related with the remaining energy capacity for discharging  $(E_0 - \underline{E})\eta^d$. In our illustrative example, point B is equals to $\min \left\{ 1, (1.5 - 0.7)0.9 \right\} = 0.72 $.

Any other points that lie within the yellow region but not within the thick black line would imply simultaneous charging and discharging. For instance, the topleft boundary of the region allows an increase of power discharge capacity (more than in point B) because a simultaneous charging creates additional room for discharging. Thus, $p_d$ is limited by $p_d \leq (E_0   - \underline{E} + \eta^c p^c)\eta^d$ instead of being limited by $p_d \leq (E_0 - \underline{E})\eta^d$. 
% Explaing 
This observation would be used to redefine a new linear BESS formulation introduced in subsection \ref{sec. new_model}.

%% Figures

\subsection{Alternative existing linear BESS models}
%%%% 

In this subsection, two additional linear BESS models are presented. The first is based on recent work by \textit{Nazir and Almassalkhi} \cite{nazir2021guaranteeing}. The second one leverages the properties of  an integer programming relaxation \cite{wolsey2020integer}. 

\textit{Nazir and Almassalkhi} \cite{nazir2021guaranteeing} provided an alternative formulation coninned by the authors  as {\textit{robust}} in the sense that it would always provide a realizable path, i.e., a feasible deployment of charging and discharging power rates while ensuring the SoE limits\footnote{It is worth to remark that the aim of \cite{nazir2021guaranteeing} is to provide a formulation that guarantees the satisfaction of the SoE constraints, while this paper does not focus on giving such guarantees. However, as novel linear BESS model  \cite{nazir2021guaranteeing}, it is deemed of particular pertinence for the scope of our paper, so we have included it in our comparative analysis.}. 
The model formulation described by \eqref{eq. Nazir} and here is referred  as \NA.

%%% Nazir LP formualtion
\begin{model}[ht!]
\caption{Nazir-Almassalkhi formulation \cite{nazir2021guaranteeing} \hfill (\NA)}
\label{Mod: robust LP}
\begin{subequations} \label{eq. Nazir}
\begin{IEEEeqnarray}{rl}
&   \overline{E} \geq  E_0 + \eta (p^c - p^d) \\
&  \underline{E} \leq  E_0 + \eta^c p^c - \frac{1}{\eta^d}p^d  \\
&  p^c + p^d \leq P^{max} \label{eq. Pc + Pd NA} \\
%
% &  e =  e_0 + \eta^c p^c - \frac{1}{\eta^d}p^d  \\
&  0 \leq  p^c  \leq \overline{P}^c \\
&  0 \leq  p^d  \leq \overline{P}^d 
% &   \underline{E} \leq  e \leq \overline{E} 
 \end{IEEEeqnarray}
 \mbox{\bf Variables: } $ p^c, p^d  \in \mathbb{R}_{\geq 0}$
\end{subequations}
\end{model}

In the \NA formulation, it is required to provide two additional parameters, a single net charging efficiency $\eta$ and a single net-charging capacity rate $P^{max}$.  When having symmetric charging and discharging capacity rates, the single net charging capacity rate can be easily selected as $P^{max} = \overline{P}^c = - \overline{P}^d$, however, this is not the case in most of the BESS. The single net-charge efficiency is not defined as the round-trip efficiency by the authors (i.e., $\eta = \eta^c \eta^d$), but it is selected such that $\eta - \eta^c = 1 - \frac{1}{\eta^d}$, therefore $ \eta = \frac{1}{2}(\frac{1}{\eta^d} + \eta^c)$. This could lead to $\eta  > 1$. 

The feasible region of the linear BESS model \NA is presented in Fig. \ref{fig: feasible regions}a in dark green, for the illustrative two-period example. In this example, the single net charging efficiency $\eta$ as described in the original article \cite{nazir2021guaranteeing}, and the single net charging capacity $P^{max} = \max \{\overline{P}^c,\overline{P}^d \}$ have been selected to not cut part of the actual feasible region. The main linear BESS formulation of the original proposal \cite{nazir2021guaranteeing} is developed without the valid inequality \eqref{eq. Pc + Pd NA}.  It was introduced in \cite{almassalkhi2014model}. Eliminating \eqref{eq. Pc + Pd NA} from \NA would result in the light green region, which approximates the feasible region \LP, Fig. \ref{fig. exact_vs_relaxed}. 
Observe that equation \eqref{eq. Pc + Pd NA} provides an important reduction of the feasible space of search where complementarity is not met.

\vspace{0.3cm}
Any integer programming model can be \textit{relaxed} to a linear programming model by eliminating integrality conditions so that the binary (integer) decision variables can take values in a continuous domain \cite{wolsey2020integer}.  The \Relyz is constructed on the \Exc formulation by relaxing the $z$ and $y$ variables in \eqref{eq. Re_LP_zy_z_rel} and  \eqref{eq. Re_LP_zy_y_rel}. The rest of the formulation of \Relyz is equal to \Exc.  

%%% Relaxed LP formualtion from MILP
\begin{model}[ht!]
\caption{Relaxed LP formulation \hfill (\Relyz)}
% \label{Mod: relased LP}
\begin{subequations} \label{eq. relaxed_zy}
\begin{IEEEeqnarray}{rl}
&  e =  E_0 + \eta^c p^c - \frac{1}{\eta^d}p^d  \\
&  0 \leq  p^c  \leq \overline{P}^c z  \label{eq. Re_LP_zy PcMax}\\
&  0 \leq  p^d  \leq \overline{P}^d y \label{eq. Re_LP_zy PdMax}\\
&   \underline{E} \leq  e \leq \overline{E}  \\
& z + y \leq 1 \label{eq. Re_LP_zy_lim} \\
& 0 \leq z \leq 1  \label{eq. Re_LP_zy_z_rel}\\
& 0 \leq y \leq 1  \label{eq. Re_LP_zy_y_rel}
\end{IEEEeqnarray}
\mbox{\bf Variables: }  $ e, p^c, p^d, z, y \in \mathbb{R}_{\geq 0} $

\end{subequations}
\end{model}

The feasible region that describes the \Relyz formulation is represented in red in Fig. \ref{fig: feasible regions}b. This formulation does not cut any feasible solution of the original model \Exc. This is a well-known property of relaxed MILP models \cite{wolsey2020integer}. The \Relyz describes the same search space as \LP, except for the cut \eqref{eq. Re_LP_zy_lim}. This is, of course, a tighter formulation close to the convex hull.  In any case, the complementarity of the charging and discharging states is not guaranteed in this formulation. 

\subsection{New linear BESS model}
\label{sec. new_model}
A new linear BESS model with additional constraints on charging and discharging power is proposed in this section. This model is referred as \ExtLP formulation. 
%This formulation is the best possible linear formulation among the proposed ones for the mixed-integer \Exc model. 
Although it does not provide the convex hull of \Exc, \ExtLP is tighter than other known BESS linear formulations.

\begin{model}[ht!]
\caption{Extended relaxed formulation \hfill (\ExtLP)}
% \label{Mod: bilenar}
\begin{subequations} \label{eq. extended_LP}
\begin{IEEEeqnarray}{rl}
&  \text{constraint set } \eqref{eq. relaxed} \text{ and  } \\
&  p^c \leq (\overline{E}-E_0)/\eta^c \label{eq. relx charg} \\
&  p^d \leq (E_0 - \underline{E})\eta^d  \label{eq. relx disch} \\
&  p^d \leq   \overline{P}^d -  (\overline{P}^d/\overline{P}^c) p^c \label{eq. pcpd_maxC} 
 \end{IEEEeqnarray}
\mbox{\bf Variables: } $ e, p^c, p^d \in \mathbb{R}_{\geq 0} $
\end{subequations}
\end{model}
The feasible region generated by the formulation \ExtLP is shown in blue in Fig. \ref{fig: feasible regions}c. 
The actual maximum charging and discharging capabilities (A and B points in Fig \ref{fig. exact_vs_relaxed}) are explicity used to construct the limits on the charging  capacity rate \eqref{eq. relx charg} and discharging capacity rate \eqref{eq. relx disch}.  
    Equation \eqref{eq. relx charg} together with \eqref{eq. LP PcMax} implies $p^c \leq  \min \left\{ \overline{P}^c, (\overline{E} - E_0)/\eta^c \right\}$. It can be seen as a tighter bound for the charging power capacity that accounts both, the rated charging power capacity and the remaining capacity at the battery for charging. Similarly, equation \eqref{eq. relx disch} together with \eqref{eq. LP PdMax} implies $p^d \leq  \min \left\{\overline{P}^d,  (E_0 - \underline{E})\eta^d \right\}$, a tighter bound for the discharging power capacity. 
Finally, equation \eqref{eq. pcpd_maxC} represents a similar, but not equivalent, cutting plane to \eqref{eq. Re_LP_zy_lim} from the \NA formulation. 
% Similar cut, but not equivalent, is given in \NA formulation \eqref{eq. Pc + Pd NA} for a symetric charing and discharging power capacity.

\subsection{Summary of linear BESS  models}

We have shown five linear BESS models, four of them suitable for linear programming (LP) implementations. Two main remarks should be done at this point. 

\subsubsection{MIP tightness}
Firstly, it is easy to see that the following relations are valid.
\begin{IEEEeqnarray}{C}
\text{\Exc~$\subseteq$~\ExtLP~$\subseteq$~\Relyz~$\subseteq$~\LP} \nonumber 
\end{IEEEeqnarray}

The \ExtLP is the tightest LP formulation from all presented models. One should expect the solutions from \ExtLP being closer to \Exc formulations. 
However, it is shown from our numerical analysis that the difference is not that important for all LP formulations, depending on the context of application. Future research could focus on using tighter formulations of the \Exc. Tight formulations are preferred in B\&B algorithms because they can provide LP relaxations that contribute to a smaller gap.

%  \item[$\bm{(i)}$]  \mbox{\NA~$\subseteq$~\LP} 

\subsubsection{Convex hull construction} Secondly, it is possible to analytically describe the convex hull in the $(p^c,p^d)$-space by using the actual maximum charging and discharging power capacity, i.e., A and B points. The convex hull is therefore constructed by constraint set \eqref{eq. relaxed} and \mbox{$p^d \leq B-(B/A)p^c$} where $A = \min \left\{ (\overline{E} - E_0)/\eta^c, \overline{P}^c \right\}$ and $B =   \min \left\{ (E_0 - \underline{E})\eta^d, \overline{P}^d \right\}$. 
Unfortunately, it leads to a nonlinear and non-convex formulation with bilinear terms.

\section{Numerical Experiments}
\label{sec. Numerical}

The five proposed BESS formulations are tested in two application contexts. We refer to a \textit{contextual problem} or just \textit{a problem} to the formal framework for the application of BESS models. Two problems are addressed here, namely, the set-point tracking  (SPT) and the transmission expansion planning (TEP) problems. 
Several random problem instances are used for the experiments. A \textit{problem instance} or just an \textit{instance} is related to a particular set of input parameters. In our case,  \textit{problem instances}  are randomly generated.

\begin{figure}[H]
     \centering
     \begin{subfigure}[b]{1\linewidth}
         \centering
         \includegraphics[width=\linewidth]{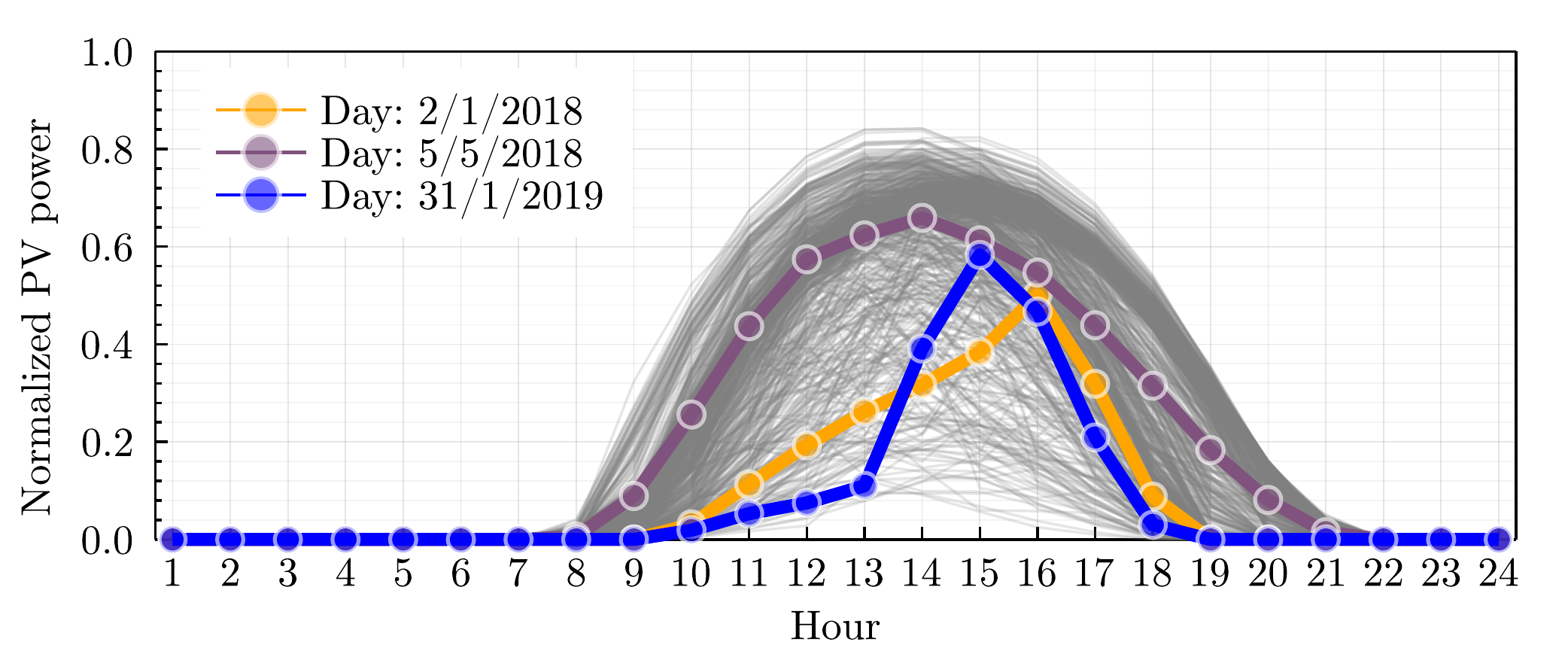}
        %  \caption{$y=x$}
        %  \label{fig:y equals x}
     \end{subfigure}
     \hfill
     \begin{subfigure}[b]{1\linewidth}
         \centering
         \includegraphics[width=\linewidth]{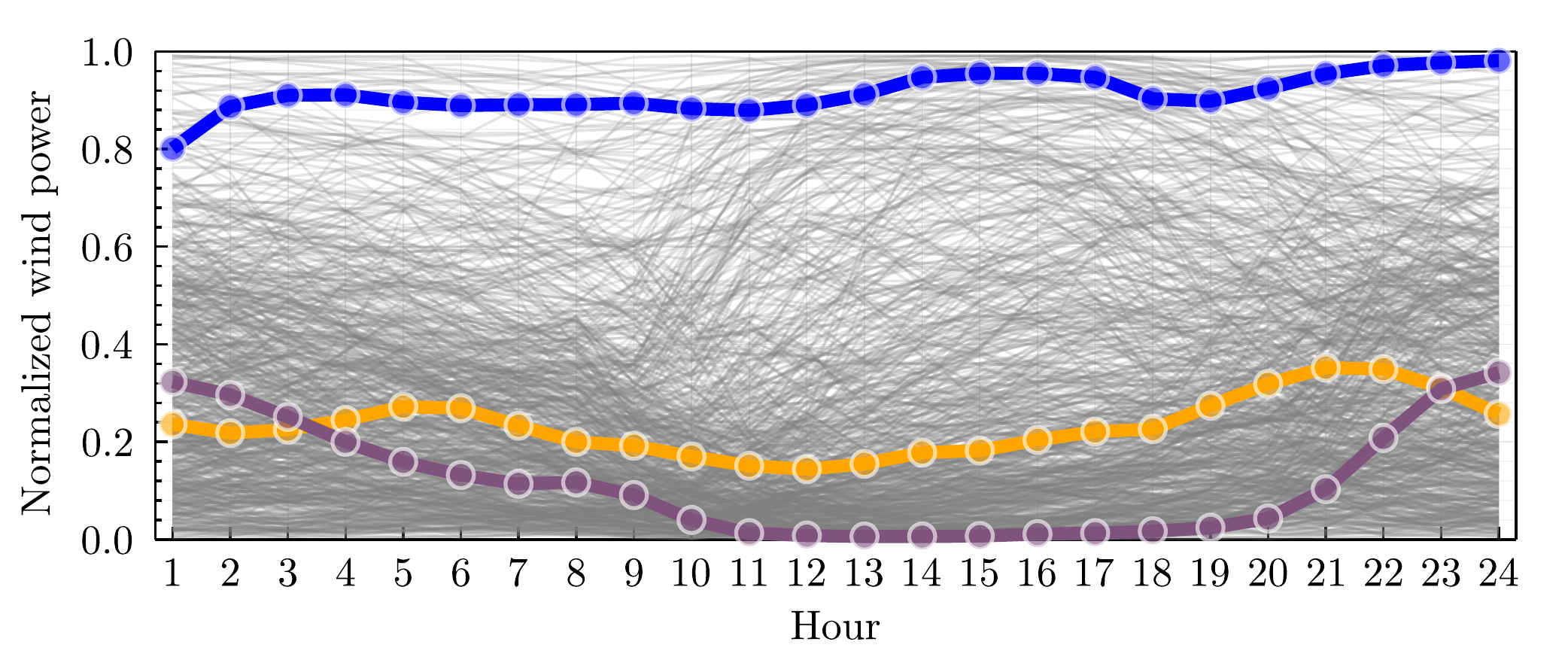}
     \end{subfigure}
        \caption{{Normalized RES daily profiles, 750 solar PV and 750  wind, sited at Castilla-La Mancha region, Spain, taken from years 2018 and 2019. Three random days are represented in orange, purple, and blue for PV and wind generation daily profiles.}}
        \label{fig: pv_wind_data}
\end{figure}

The data used for each problem instance has been synthetically generated. It is  available in \cite{data_source}.  
1450 daily profiles for wind and solar generation were generated from 2018 and 2019 using \textit{renewables.ninja} \cite{pfenninger2016long}. Fig \ref{fig: pv_wind_data} sumarizes the normalized profiles for a unit-capacity power plant. One hundred BESS were randomly generated using the parameters listed in Table \ref{tab. bess parameters random}. The initial state of BESS is set to $E_0 = (\overline{E} + \underline{E})/2$ for 100 random BESS. 

\begin{table}[!th]
\renewcommand{\arraystretch}{1.3} % adds row cushion
\addtolength{\tabcolsep}{-2pt} 
\caption{BESS parameter generation}
\centering
% \resizebox{0.4\paperwidth}{!}{
\begin{tabular}{|cc|cc|cc|}
  \hline      
    $\mathbf{\underline{E}}$  & $\mathbf{\overline{E}}$ &  $\mathbf{\overline{P}^c}$  & $\mathbf{\overline{P}^d}$  & $\mathbf{\eta^c}$ & $\mathbf{\eta^d}$ \\
\hline      
  \hline
$\unif(0,30)$ & $\unif(40,80)$ & $\unif(10,20)$ & $\unif(10,20)$ & $\unif(0.75,1)$ & $\unif(0.75,1)$ \\ 
  \hline
\end{tabular}
% } 
\vspace{0.3cm}
\footnotesize{$\unif(a,b)$ represents a continuous Uniform distribution between $a$ and $b$.} \label{tab. bess parameters random}
\end{table}

All numerical experiments were run to optimality by using the branch-and-cut algorithm of Gurobi 9.2 and default options (optimality gap equals to $10^{-3}$). The implementation was carried out on Julia v1.6.3 \cite{Julia} with JuMP v0.21.6 \cite{JuMP}, and was executed on an iMac with 3.4 GHz Quad-Core Intel Core i5, 64 GB of RAM.
%except for the optimality gap that were set to 0\%. 

\subsection{Set-point tracking problem}

The first context of application is the set-point tracking problem (SPT). It is assumed that an aggregator decides the control actions for $N$ BESS (charging and discharging) at each time $t$ such as a net power demand signal tracked by the use of BESS units \cite{alahyari2021online}. In other words, the SPT problem aims to match a net power demand (actual demand minus local renewable production) by BESS.

The SPT is formulated as the minimization of the squared tracking error between the control actions of $N$ BESS and a power signal to follow $p_t^{\text{sig}}$. The selection of this problem is motivated by its use in the numerical analysis in \cite{nazir2021guaranteeing}, one of the presented BESS models.  The SPT problem \eqref{eq. SPT} belongs to the class of quadratic programming problems with linear constraints.
\begin{subequations} \label{eq. SPT}
% \vspace{-0.5cm}
\begin{IEEEeqnarray}{ll}
\text{\bf{(SPT)}}  \quad & \min  \: \:   \sum_{t \in T} \left(  \sum_{n \in N}\left( p^d_{nt} - p^c_{nt} \right) - p_t^{\text{sig}} \right)^2 \\
\vspace{0.5cm}
& \text{s.t.:}  \: \: \text{BESS model }(n,t), \: \forall n \in N, t \in T 
\end{IEEEeqnarray}
\end{subequations}
The horizon is set to 24 hours with an interval period of 1 hour.
One hundred instances of the SPT problem are run. Each instance is generated by selecting a BESS and RES profile randomly.  
Each of the 100 random instances of the SPT has the same input parameters for the 5 BESS formulations.
The experiment is repeated for different numbers of BESS ranging from 1 to 5 to examine scalability.  

\begin{figure}[bht]
\centering
    % \includesvg[width=1\linewidth]{figs/exmple_SPTP.svg}
    \includegraphics[width=\linewidth]{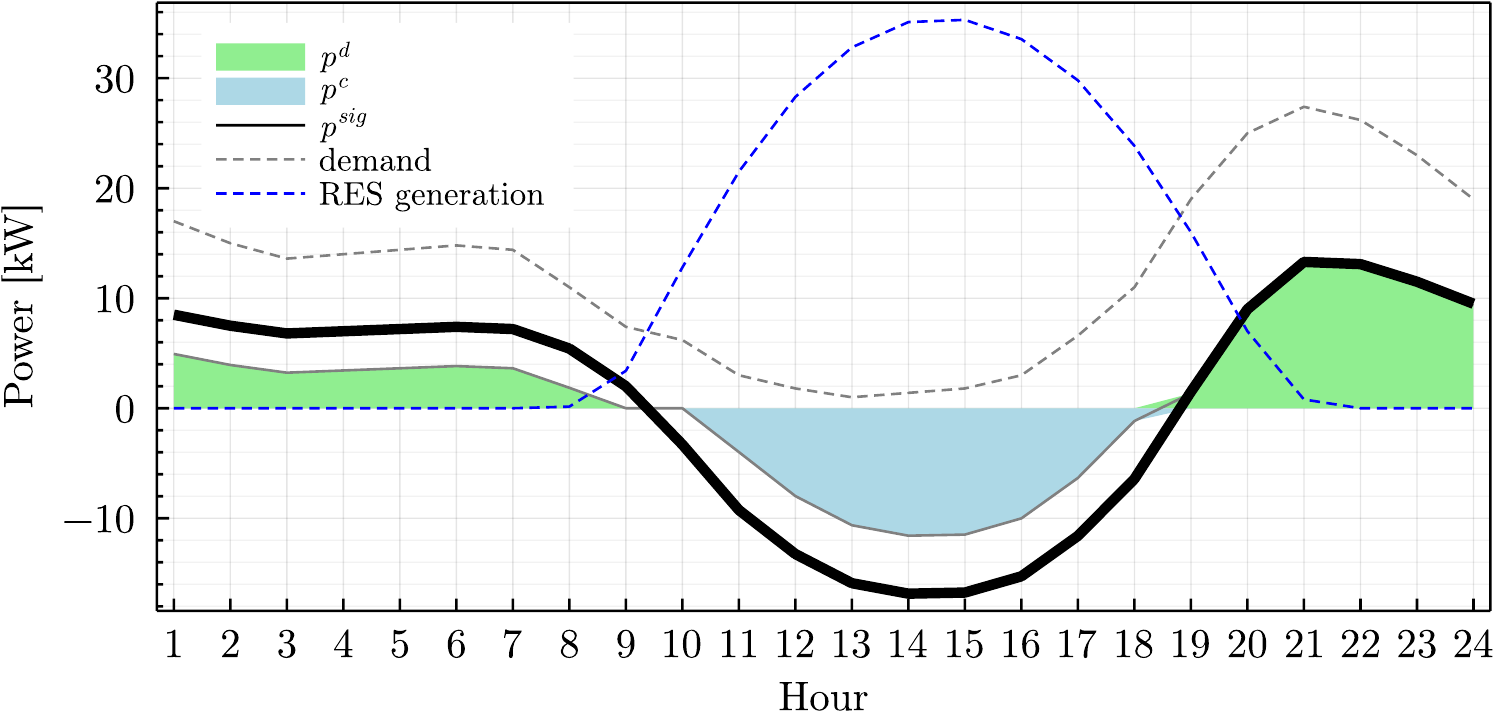}
	\caption{Data and solution example for the SPT problem and a single instance when using \Exc BESS formulation and one BESS unit. BESS parameters are randomly selected. The $p^{\text{sig}}$ is build upon a deterministic demand and a random RES profile. $p^c$  and $p^d$ are solutions of \eqref{eq. SPT}.}
	\vspace{0.5cm}
	\label{fig. example_sptp}
\end{figure}

The BESS models are compared in RMSE value, events with simultaneous charging and discharging (materialized by the condition $|p_t^c p_t^d| > 10^{-4}$), and solution time.  
Fig. \ref{fig. example_sptp} shows the data and optimal results for one problem instance. A deterministic demand profile and a random RES generation profile are used for generating the net demand, i.e., the signal $p^{\text{sig}}$ to follow (black line)  with the available pool of BESS. Solution of the charging and discharging for this particular instance are also represented in blue and green, respectively. 

A summary of the results is given in Table \ref{tab. SRP summary}. The first column indicates the number of BESS used, while the second column refers to the BESS model. The third column summarizes the frequency at which charging and discharging occurred simultaneously. Observe that no LP model avoids simultaneous charging and discharging with a large percentage in all cases (around 30\%). However, the \ExtLP is slightly better in this aspect. The fourth column represents the average resolution time of each problem instance. As expected for linear models, the solution time grows polynomially (linearly) with the number of BESS, while this is not the case for the  \Exc formulation. The last column represents the RMSE relative to the \Exc formulation. All the LP models for all experiments provide smaller RMSE values except for 3 cases of the \NA formulation. However, it is not fundamentally different. An interesting case is the \LP formulation for $N=1$ where the RMSE is 15\% lower than the \Exc. This highlights the overoptimistic behavior of the \LP formulation in the context of the SPT problem.
That is, control signals can fit better to the signal to follow because of the simultaneous charging and discharging that happens when there is an excess of renewable generation. 

\begin{table}[h!]  
\renewcommand{\arraystretch}{1.2} % adds row cushion
% \addtolength{\tabcolsep}{-2pt} 
\centering
\caption{Summary of numerical experiments for 100 instances of the SPT problem . The values presented are the averages of all 100 instances.}
\begin{tabular}{|c|c|c|c|c|}
\hline
\multicolumn{1}{|c|}{\textbf{N}} &
  \multicolumn{1}{c|}{\textbf{BESS Model}} &
  \multicolumn{1}{c|}{ $|p_t^c p_t^d| > 10^{-4}$ \textbf{[\%]}} &
  \multicolumn{1}{c|}{\textbf{Time [s]}} &
  \multicolumn{1}{c|}{\textbf{RMSE [\%]}} \\ 
  \hline
  \hline
1 & \multirow{5}{*}{\Exc} & 0.0   & 0.2148 & 1.0  \\
2 &  & 0.0   & 0.0617 & 1.0  \\
3 &  & 0.0   & 0.0493 & 1.0  \\
4 &  & 0.0   & 1.9792 & 1.0  \\
5 &  & 0.0   & 1.2353 & 1.0  \\
  \hline
  \hline
1 & \multirow{5}{*}{\LP} & 40.42 & 0.0008 & 0.85 \\
2 &  & 39.16 & 0.0011 & 0.91 \\
3 &  & 26.25 & 0.0015 & 0.93 \\
4 &  & 43.75 & 0.0017 & 0.87 \\
5 &  & 24.58 & 0.0024 & 0.92 \\
  \hline
  \hline
1 & \multirow{5}{*}{\NA} & 39.58 & 0.0010  & 0.96 \\
2 &  & 32.5  & 0.0020  & 1.0  \\
3 &  & 24.45 & 0.0024 & 0.99 \\
4 &  & 39.37 & 0.0030  & 1.02 \\
5 &  & 21.25 & 0.0037 & 1.03 \\
  \hline
  \hline
1 & \multirow{5}{*}{\Relyz} & 40.42 & 0.0010  & 0.91 \\
2 &  & 39.16 & 0.0020  & 0.95 \\
3 &  & 26.25 & 0.0027 & 0.97 \\
4 &  & 43.75 & 0.0032 & 0.96 \\
5 &  & 24.58 & 0.0047 & 0.97 \\
  \hline
  \hline
1 & \multirow{5}{*}{\ExtLP} & 35.42 & 0.0011 & 0.93 \\
2 &  & 25.83 & 0.0022 & 0.98 \\
3 &  & 14.31 & 0.0030  & 0.99 \\
4 &  & 32.4  & 0.0036 & 0.99 \\
5 &  & 18.92 & 0.0047 & 0.99 \\
  \hline
\end{tabular} \label{tab. SRP summary}
\end{table}

\subsection{Transmission expansion planning problem}

The following application context is the transmission expansion planning (TEP) problem. In the TEP problem, new updates of the grid are optimally selected by considering the future operation of the grid for the newly updated topology \eqref{eq. TEP}. The TEP problem minimizes capital expenditures, CAPEX, related to the investment in new lines, and operational expenses, OPEX, associated with the operating cost over several years. Further details of a standard centralized TEP formulation can be found in \cite{pozo2017doing}. Conceptually, the TEP problem is formulated as follows.
\begin{subequations} \label{eq. TEP}
% \vspace{-0.5cm}
\begin{IEEEeqnarray}{rrl} 
\text{\bf{(TEP)}} \quad  &  \; \: \min  \; \:&  \Big(  \text{CAPEX} + \text{OPEX} \Big)\\
% \vspace{0.5cm}
 &\text{s.t.:} \; \:  &   \text{Investment restrictions} \\ 
 &  \: \: & \text{Network flow equations}  \\
 & \: \: &  \text{Network capacity limits}  \\
 &\: \: &  \text{Generator limits}  \\
 &\: \: &  \text{BESS model.}  \quad
\end{IEEEeqnarray}
\end{subequations}

\begin{figure}[bht]
\centering
    \includegraphics[width=0.6\linewidth]{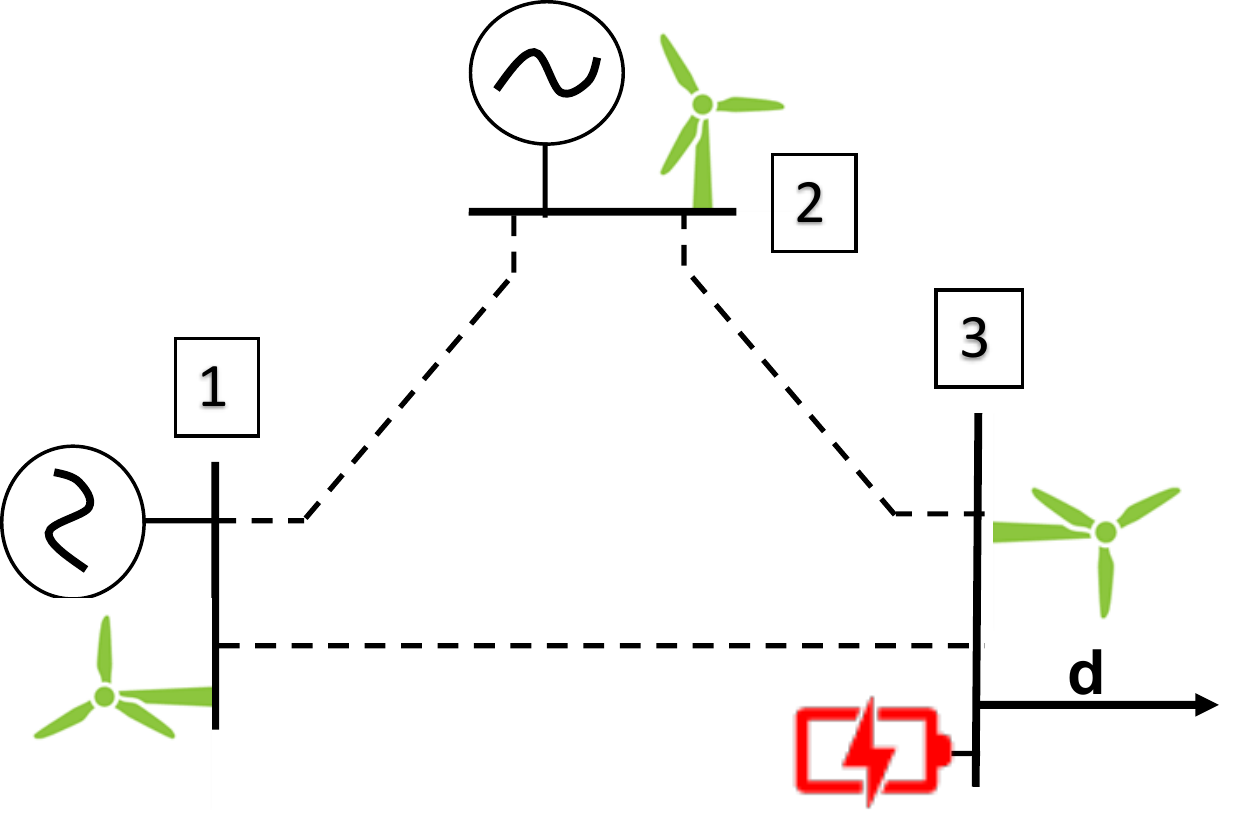}
	\caption{Three-node system used for experiments. Each node has wind renewable generation randomly selected from the dataset \cite{data_source}. Demand and storage devices are located at node 3 while node 1 and 2 have conventional generators.}
	\vspace{0.5cm}
	\label{fig. 3nodesystem}
\end{figure}

An illustrative problem of three nodes, depicted in Fig. \ref{fig. 3nodesystem} is used for validation. The OPEX comprises the cost of 50 typical days. Thus, every TEP instance has 50 randomly selected daily wind profiles for each node and $N$ random BESS units located at node 3.  The experiment is repeated for different numbers of BESS ranging from 1 to 5 to examine scalability, and 100 instances for each case. Therefore, each MILP instance that uses model \eqref{eq. exact} would have $2 \times 24 \times 50 \times N$ binary variables.

Table \ref{tab. TEP summary} compares the performance of each BESS formulation for a different number of BESS units available in the TEP problem. The table summarizes the 100 solved instances. Similarly to the Table \ref{tab. SRP summary}, percentage of simultaneous charging and discharging events, and the average time of resolution are reported. It can be seen that in this TEP problem, simultaneous charging and discharging events rarely occur for all LP formulations.
Interestingly, the total cost (optimal value from \eqref{eq. TEP}) is almost the same for all BESS models and the number of BESS considered. This is due to the optimal investment in lines for all BESS models being nearly the same. It demonstrates the practical use of LP BESS formulations in the context of the TEP problem. The optimal value errors ($\approx 0$\%) and solutions errors ($\approx 1-3$\%) are irrelevant for practical decision-making for the long-term planning. 

\begin{table*}[h!t]
\renewcommand{\arraystretch}{1.2} % adds row cushion
\addtolength{\tabcolsep}{-2pt} 
\centering
\caption{Summary of numerical experiments for 100 instances of TEP problems. The values presented are the averages of all 100 instances.}
\begin{tabular}{|c|c|c|c|c|c|c|c|c|}
  \hline
\multicolumn{1}{|c|}{\textbf{N}} &
  \multicolumn{1}{c|}{\textbf{BESS Model}} &
  \multicolumn{1}{c|}{ $|p_t^c p_t^d| > 10^{-4}$ \textbf{[\%]}} &
  \multicolumn{1}{c|}{\textbf{Time [s]}} &
  \multicolumn{1}{c|}{\textbf{Total rel. cost [\%]}} &
  \multicolumn{1}{c|}{\textbf{Load shedding [pu]}} & 
  \multicolumn{1}{c|}{\textbf{RES curtailment [pu]}} &
    \multicolumn{1}{c|}{\textbf{Total capacity invs. [pu]}} 
  \\ 
  \hline
  \hline
1 & \multirow{5}{*}{\Exc} & 0.0  & 0.3265 & 1.0  & 0.0 & 4.567 & 59.939 \\
2 &                    & 0.0  & 0.6092 & 1.0  & 0.0 & 2.087 & 60.163 \\
3 &                    & 0.0  & 0.3808 & 1.0  & 0.0 & 0.445 & 59.689 \\
4 &                    & 0.0  & 0.3562 & 1.0  & 0.0 & 0.667 & 60.01  \\
5 &                    & 0.0  & 0.2883 & 1.0  & 0.0 & 0.483 & 60.101 \\
  \hline
  \hline
  1 & \multirow{5}{*}{\LP} & 3.28 & 0.0142 & 0.99 & 0.0 & 0.536 & 59.939 \\
2 &                    & 1.46 & 0.0207 & 0.99 & 0.0 & 0.197 & 60.163 \\
3 &                    & 0.43 & 0.0288 & 1.0  & 0.0 & 0.073 & 59.687 \\
4 &                    & 0.65 & 0.033  & 1.0  & 0.0 & 0.091 & 60.01  \\
5 &                    & 0.44 & 0.0418 & 1.0  & 0.0 & 0.075 & 60.101 \\
  \hline
  \hline
  1 & \multirow{5}{*}{\NA} & 2.48 & 0.0157 & 1.0  & 0.0 & 3.432 & 59.939 \\
2 &                    & 1.58 & 0.0255 & 1.0  & 0.0 & 1.731 & 60.148 \\
3 &                    & 0.63 & 0.037  & 1.0  & 0.0 & 0.421 & 59.688 \\
4 &                    & 1.13 & 0.0474 & 1.0  & 0.0 & 0.613 & 60.011 \\
5 &                    & 0.86 & 0.0584 & 1.0  & 0.0 & 0.375 & 60.101 \\
  \hline
  \hline
  1 & \multirow{5}{*}{\Relyz} & 4.06 & 0.0175 & 0.99 & 0.0 & 1.458 & 59.94  \\
2 &                    & 1.95 & 0.025  & 1.0  & 0.0 & 0.702 & 60.163 \\
3 &                    & 0.58 & 0.0357 & 1.0  & 0.0 & 0.146 & 59.69  \\
4 &                    & 0.93 & 0.0489 & 1.0  & 0.0 & 0.22  & 60.01  \\
5 &                    & 0.67 & 0.0686 & 1.0  & 0.0 & 0.148 & 60.101 \\
  \hline
  \hline
  1 & \multirow{5}{*}{\ExtLP} & 3.32 & 0.0165 & 0.99 & 0.0 & 2.773 & 59.939 \\
2 &                    & 1.43 & 0.0243 & 1.0  & 0.0 & 1.389 & 60.163 \\
3 &                    & 0.49 & 0.0345 & 1.0  & 0.0 & 0.284 & 59.689 \\
4 &                    & 0.66 & 0.0434 & 1.0  & 0.0 & 0.472 & 60.01  \\
5 &                    & 0.45 & 0.0563 & 1.0  & 0.0 & 0.359 & 60.101 \\
  \hline
\end{tabular} \label{tab. TEP summary}
\end{table*}

\subsection{Computational performance curves}

\begin{figure*}[h!]
     \centering
     \begin{subfigure}[b]{0.48\linewidth}
         \centering
         \includegraphics[width=\linewidth]{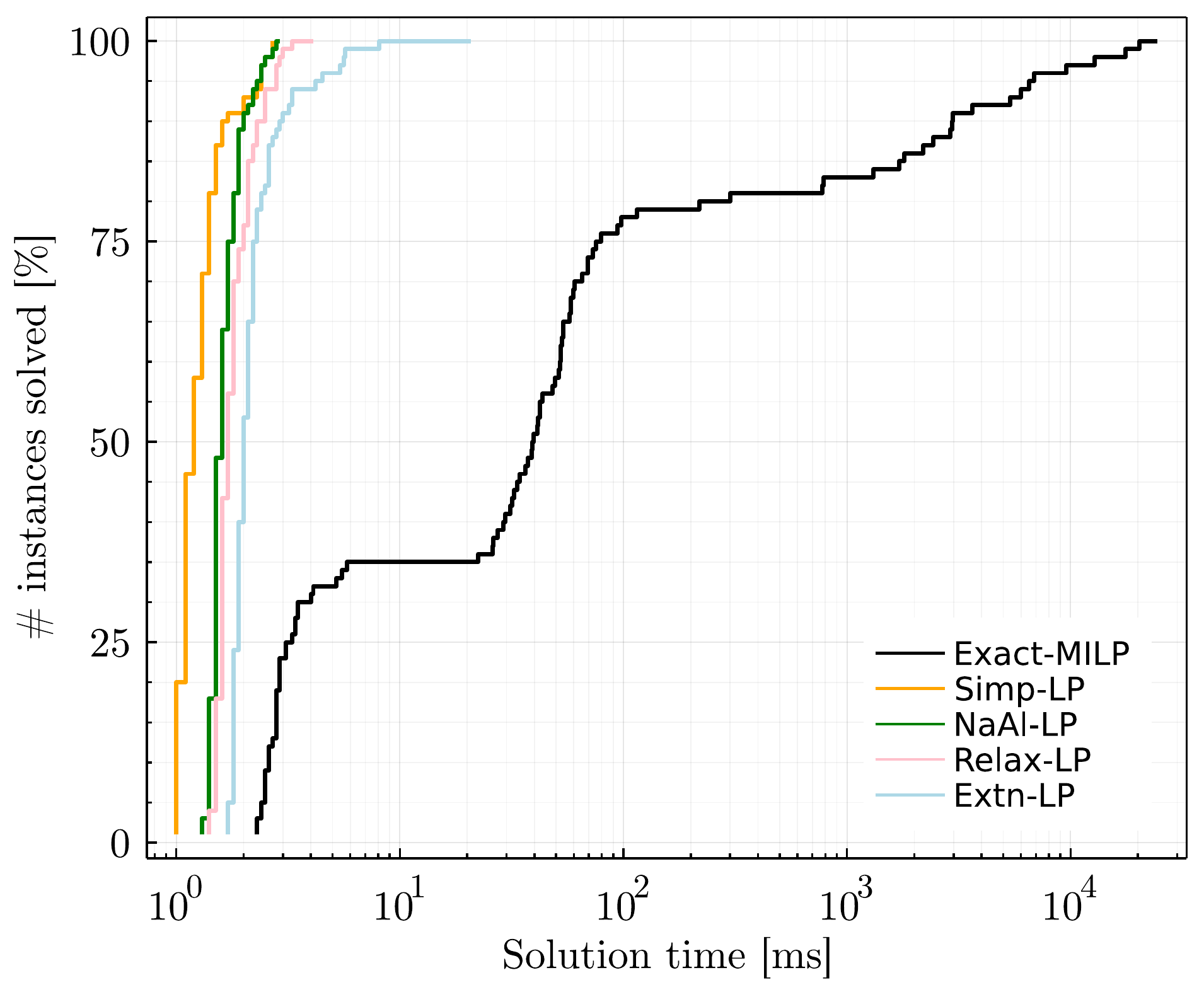}
        %  \caption{$y=x$}
        %  \label{fig:y equals x}
     \end{subfigure}
     \hfill
     \begin{subfigure}[b]{0.48\linewidth}
         \centering
         \includegraphics[width=\linewidth]{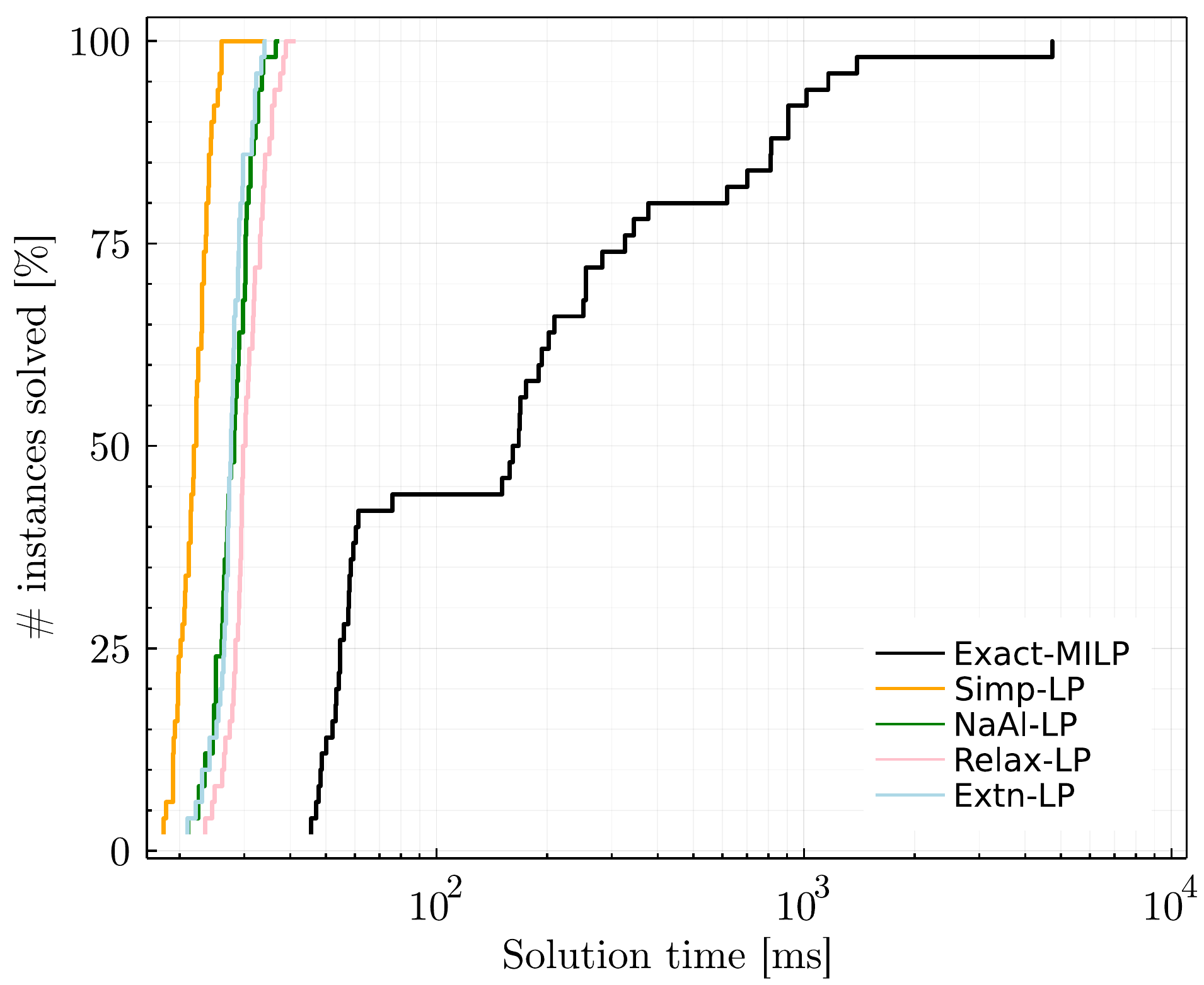}
     \end{subfigure}
        \caption{Computational performance curves representing percentage of instances solved vs. runtime for solving each of them. \textit{Left:} SPT problem with two BESS units.  \textit{Right:} TEP problem with 50 typical days.}
        \label{fig: performance results}
\end{figure*}

In Fig. \ref{fig: performance results},  performance curves are shown for each BESS model and the two contextual problems, SPT and TEP. The y-axis represents the percentage of instances solved, while the x-axis (in logarithmic scale) corresponds to the total time for solving each of the 100 random problem instances in milliseconds. 
Although average runtime can be an attractive indicator for comparing BESS formulation performance, it does not fully describe the computational performance of each BESS formulation. 
For example, it can be observed that an instance of \Exc formulation for the SPT problem can be solved in a couple of milliseconds or multiple tens of thousands of milliseconds, that is, four orders of magnitude difference. A comparison of a simple instance could be as good as the LP formulations or as bad as the worst observed case (4 orders of magnitude).    

In both problem contexts, all LP BESS formulations performed quite similarly. The \LP is slightly better and the \ExtLP is slightly worst that others. However, this difference is not significant. Undoubtedly, the MILP formulation is the worst in terms of computational performance, but at the same time, this BESS formulation showed notable variability. Thus, conclusions from a single problem instance, a standard benchmark analysis in many academic works, could lead to misidentification of practical BESS formulations.

\section{Conclusions}
\label{sec. conclusions}

In this paper, five linear BESS formulations are presented. Four of them have already been reported, and a new one is proposed. The new linear BESS formulation is proved to be a tighter approach than the others. 
Linear BESS formulations are the cornerstone of larger operational and planning models that consider storage systems. 
Two contexts of application, set-point tracking, and transmission expansion planning problems, are used to test the formulations. In total, 5000 random problem instances are solved to draw conclusions on practical aspects such as accuracy and the long-standing discussion of simultaneous charging and discharging state. All LP formulations were shown to perform very similarly in terms of computational runtime and accuracy. LP-based BESS formulation accuracy was comparable to the MILP formulation for the transmission expansion planning problem. However, no LP BESS formulation produces a decent accuracy comparison with respect to the MILP formulation in the set-point tracking problem. 

As the main takeaway, it has been seen that the performance of BESS formulations is entangled in their contextual application problem. A simple standard linear BESS model could provide competitive accuracy with a resolution time of several orders of magnitude. Further research is needed in the direction of increasing the pool of contextual problems of application.

\section*{Acknowledgments}
This paper was prepared as a part of the AMPaC Megagrant project supported by Skoltech and The Ministry of Education and Science of Russian Federation, Grant Agreement No 075-10-2021-067, Grant identification code 000000S707521QJX0002. 
This project was partially funded by the Skolkovo Institute of Science and Technology as part of the Skoltech NGP Program (Skoltech-MIT joint project).

%%%%%%%%%%%%%%%%%%%%%%%%%%%%%%
%%      References          %%
%%%%%%%%%%%%%%%%%%%%%%%%%%%%%%
\bibliographystyle{ieeetr}
\bibliography{refs}

\end{document}